\documentclass{appolb}
\usepackage{graphicx}

\begin{document}
\title{Flow anisotropies due to momentum deposition from hard partons%
\thanks{Presented at Excited QCD, Tatransk\'a Lomnica, Slovakia, March 8--14, 2015}%
}
\author{Boris Tom\'a\v{s}ik$^{a,b}$, Martin Schulc$^{b,c}$
\address{$^a$ Univerzita Mateja Bela, Bansk\'a Bystrica, Slovakia \\
$^b$ FNSPE, Czech Technical University in Prague, Prague, Czech Republic\\
$^c$ \'UJV \v{R}e\v{z} a.s., \v{R}e\v{z} u Prahy, Czech Republic}}
\maketitle
\begin{abstract}
In nuclear collisions at the LHC large number of hard partons are created in initial 
partonic interactions, so that it is reasonable to suppose that they do not thermalise
immediately but deposit their energy and momentum later into the evolving 
hot quark-gluon fluid. We show that this mechanism leads to contribution to 
flow anisotropies at all orders which are non-negligible and should be taken 
into account in realistic simulations.
\end{abstract}
  
Hot and dense strongly interacting matter which is produced in ultrarelativistic
nuclear collisions like those at the LHC or RHIC, expands swiftly and soon decouples into 
individual hadrons. Ultimately, one wants to investigate its evolution since
it potentially bears an imprint of the Equation of State (EoS) and transport coefficients, 
which one wants to infer.

When the distribution of produced hadrons in transverse plane is studied, it appears
far from isotropic even in most central collisions. Customarily, one studies its 
Fourier decomposition in azimuthal angle 
\begin{equation}
\frac{dN}{p_tdp_td\phi} = \frac{dN}{p_t dp_t} \frac{1}{2\pi} 
\left [  1 + \sum_{n = 1}^\infty 2 v_n \cos \left ( n ( \phi - \psi_n) \right ) \right ]
\label{e:aflow}
\end{equation}
where $v_n$ and $\psi_n$ are parameters. 
Presently available statistics allows to determine the first six terms of such series even for 
individual events. 

The anisotropy is caused by  anisotropic expansion of the hot fireball. Due to transverse 
expansion, transverse momentum spectra of hadrons are blue-shifted. This is because 
particles are produced by regions of the fireball which move transversely outwards 
in the direction towards the detector. Waves emitted by a source moving towards 
the detector are recorded with shorter wavelength than emitted in the rest frame 
of the source. When translated in terms of momenta, this leads to 
an enhancement of higher momenta.
This mechanism connects the transverse expansion of the fireball with the shape of the 
transverse momentum spectrum and maps the anisotropies of the former onto the 
anisotropies of the latter. The largest among the anisotropies is the so-called elliptic flow. 
It is mainly caused by the second-order anisotropy in the expansion velocity in 
non-central collisions due to initial anisotropic geometry of the fireball. 

A classical argument then connects direction dependent transverse expansion velocities
with inhomogeneities in pressure distributions within the fireball. Distribution 
of the energy density is established after the incoming partons interact and shows 
quantum fluctuations.  This is in addition to any geometrical anisotropies which are due to 
non-zero impact parameter in the collision. The evolution from such an initial state is 
reasonably well described by hydrodynamic models. This is the point where 
EoS and transport coefficients enter the game. The resulting state of the expanded
and cooled fireball depends on EoS and transport properties. Thus by calculating hadronic 
spectra and their anisotropy one hopes to be able to tune e.g.\ the viscosities 
until an agreement with data is reached. 

There is a caveat, however, in this game. The initial conditions are not measured and 
can only be determined in various models. Unfortunately, they influence strongly
the resulting anisotropies of the hadron distributions \cite{luzum}. 
Thus without the knowledge 
of the initial conditions the intended strategy for the extraction of transport properties 
seems jeopardised. Lucky enough, simulations with both ideal and viscous 
hydrodynamic models show that there is quite a linear relation between the initial state 
spatial anisotropy and the anisotropy of hadron distribution \cite{Niemi,Schenke,Larry_here}.
This is true for the second and third orders and breaks for higher orders \cite{Niemi}.
That can be understood, since higher order terms are influenced by the interference 
of lower orders (so that e.g.\ $v_4$ gets contribution from the square of 
$\varepsilon_2$---second-order
spatial distribution anisotropy). The linear relation allows to map the event-by-event
fluctuations of $v_2$ and $v_3$ onto the fluctuations of $\varepsilon_2$ and $\varepsilon_3$
and thus identify the model for the initial state which best agrees with the data. 

Usually, it is assumed that there is no contribution to fluctuations during the 
hydrodynamic evolution. This may not be the case, however. Hydrodynamic simulation
with the fluid energy and momentum density coupled to dynamically fluctuating 
order parameter field shows fluctuations in the energy density which can well cause 
flow anisotropies observable in data \cite{herold}. 
More precise quantitative impact of such a mechanism 
remains to be studied. 

Here, we introduce another mechanism that can induce flow anisotropies. 
Hard partons from initial scatterings do not thermalise immediately as they are produced, 
but fly into the hot and dense quark-gluon plasma. There, mostly they are  fully quenched
so that their momentum is transferred into the fluid and must show up in the flow 
pattern. Directions of these partons are distributed isotropically, but due to their finite 
number there may be a contribution to flow anisotropy.

In addition to that, the energy in a collision at the LHC is so large that there may be pairs 
of hard partons which are close in rapidity and are directed so that they might 
come close to each other during the evolution of the collision. Even if they are 
fully quenched  before they could actually meet, their momentum has been shown 
to be further carried by the generated streams in the fluid \cite{Betz}. 
There is a good chance 
that such streams merge into one. In peripheral collisions, through this mechanism 
the collective flow induced by hard partons tends to be directed in the reaction plane
(which is spanned by the beam and the direction of the impact parameter). 

\begin{figure}[h!]
\centerline{\includegraphics[width=0.26\textwidth]{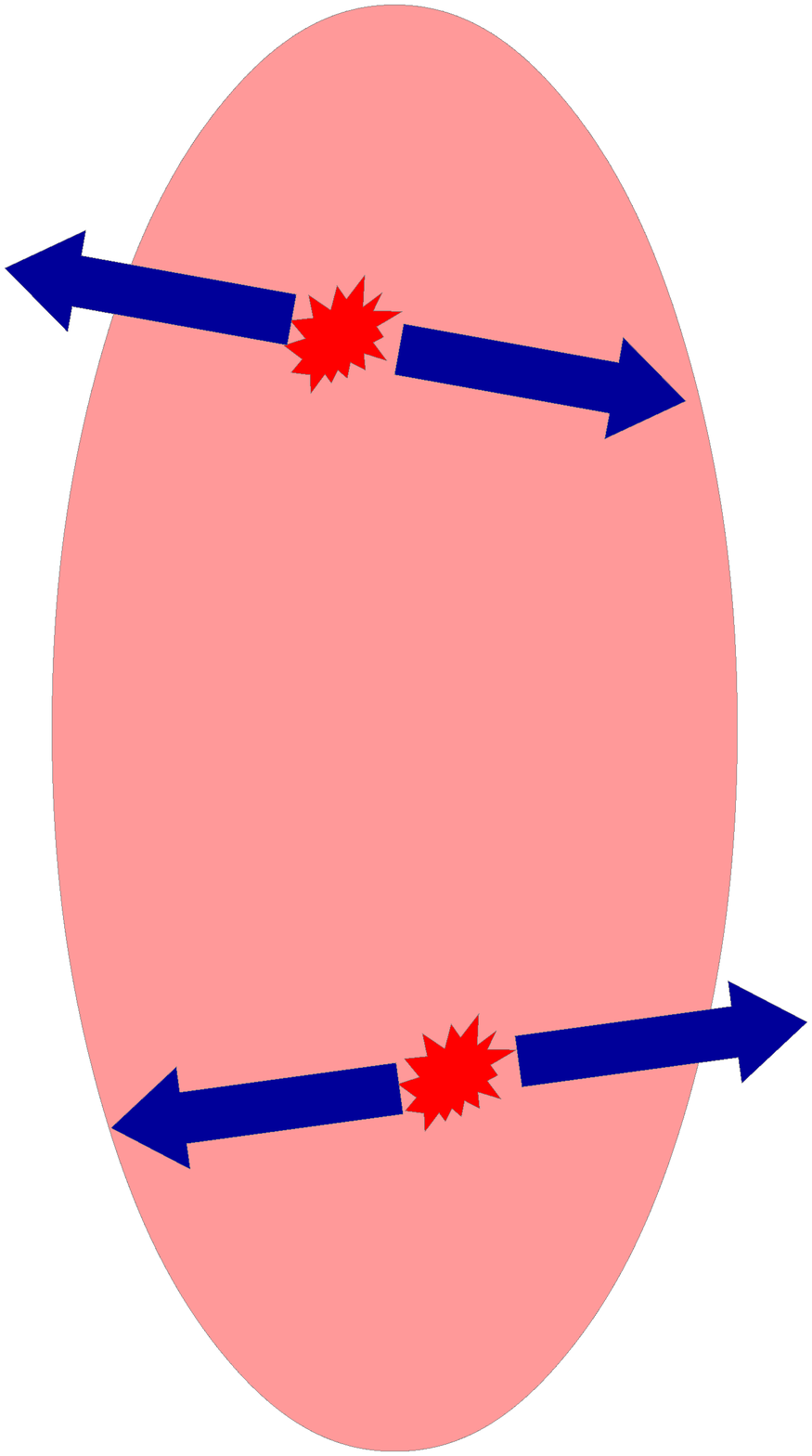} %
\includegraphics[width=0.26\textwidth]{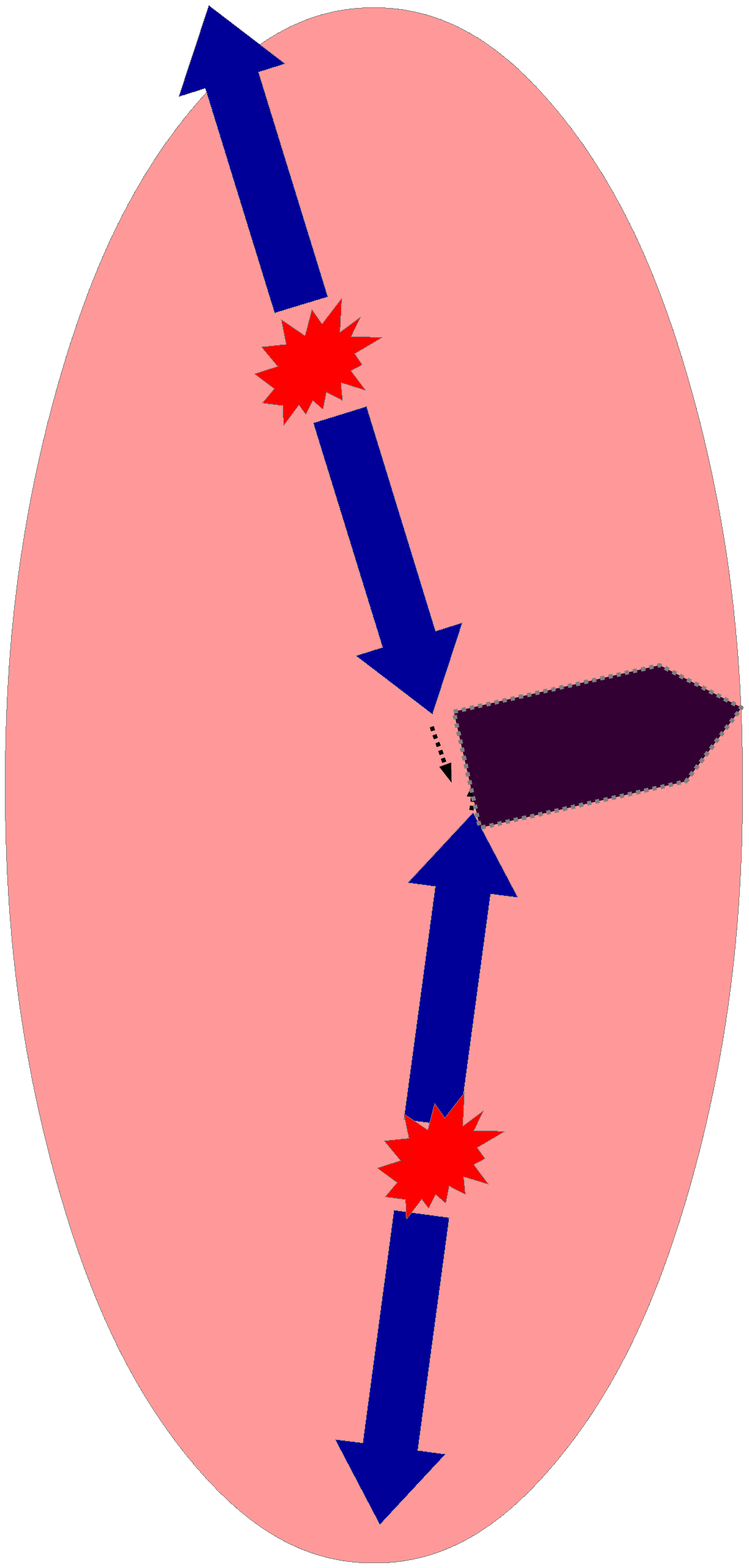}}
\caption{\label{f:cartoon}%
Transverse cross-section of the fireball in non-central collisions, 
with two dijet pairs produced. Reaction plane 
is horizontal.
}
\end{figure}

Imagine (Fig.~\ref{f:cartoon})
the almond-shaped cross-sectional area of a fireball in non-central collisions 
with two pairs of hard partons produced. If all four partons are directed parallel to 
the reaction plane, through the momentum loss they will all positively contribute to the 
elliptic flow (Fig.~\ref{f:cartoon} left). If, however, they are directed perpendicularly to the reaction plane, 
there is a chance that two of the streams will meet, merge and cancel 
flow in that direction (Fig.~\ref{f:cartoon} right). The chance is bigger in 
this direction because here the fireball is narrower and the streams have less space to 
avoid each other. 

We have set up a three-dimensional ideal hydrodynamic model and tested this 
scenario \cite{schulc1,schulc2}.
Note that a 3D simulation in this case is mandatory, since the presence of hard partons
breaks the boost-invariance (and azimuthal symmetry). 
In order to esimate the effect on flow anisotropies due to hard partons we formulated the 
model without any other fluctuations in the initial conditions. Thus our initial energy 
density profile was determined from an optical Glauber model and the distribution in 
transverse plane was determined from a combination of wounded-nucleon
and binary-collision density. The energy density in the central cell at the initial 
time of $\tau_0 = 0.5$~fm/$c$ was set to 60~GeV/fm$^3$. The initial profile 
in rapidity is flat over 10 units with Gaussian tails at the edges. 

To implement the effect of energy and momentum deposition from hard partons into plasma,
terms $J^\nu$ which represent forces are added into the energy and momentum conservation 
equation 
\begin{equation}
\partial_\mu T^{\mu\nu} = J^\nu
\end{equation}
where $T^{\mu\nu}$ is energy-momentum tensor. The force field is parametrised with 
the help of Gaussians \cite{Betz,schulc1}
\begin{equation}
J^\nu = \sum_i \frac{1}{(2\, \pi\, \sigma_i^2)^{\frac{3}{2}}} \, \exp \left (
- \frac{\left ( \vec x - \vec x_{\mathrm{jet},i} \right )^2 }{2\, \sigma_i^2} \right )\, 
\left ( \frac{dE_i}{dt}, \frac{d\vec P_i}{dt} \right )
\end{equation}
where the terms in the bracket stand for the rate of energy and momentum deposition 
and the Gaussian ($\sigma = 0.3$~fm) distributes it around the trajectory of the parton until 
all its energy is used up. The actual energy loss scales with the entropy density as 
\begin{equation}
\frac{dE}{dx} = \left . \frac{dE}{dx}\right |_0 \frac{s}{s_0}
\end{equation}
with $s_0 = 78.2/\mathrm{fm}^3$ (corresponding to $\varepsilon_0 = 20$~GeV/fm$^3$)
and $\left . dE/dx\right |_0$ being a parameter that we tuned. The places of hard parton 
production are distributed according to binary collision density in the transverse plane and 
uniformly in rapidity. Their directions are azimuthally symmetric and the $p_t$ spectrum 
follows from \cite{Levai}
\begin{equation}
\frac{1}{2\pi}\, \frac{1}{p_t}\, \frac{d\sigma_{NN}}{dp_t\, dy}
= \frac{B}{(1+p_t/p_0)^n}\,  .
\end{equation}
with $B = 14.7$~mb/GeV$^2$, $p_0 = 6$~GeV, and $n = 9.5$.

\begin{figure}[ht]
\includegraphics[width=0.98\textwidth]{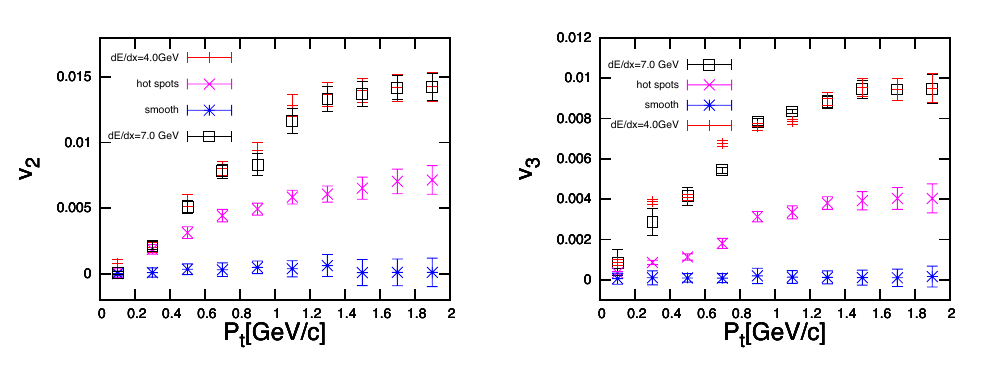}
\caption{%
Coefficients $v_2$ and $v_3$ from central collisions. Data 
calculated with two different values of hard parton
energy loss, one simulation with only energy and no momentum deposition 
(hot spots). Shown also reference simulation with smooth initial conditions. 
}%
\label{f:central}
\end{figure}

In Fig.~\ref{f:central} we show results obtained for collisions at vanishing impact parameter.
There is no anisotropy in case of no hard partons as we initiate the simulation with
smooth azimuthally symmetric energy density profile. Momentum deposition 
\emph{during} the evolution leads to measurable flow anisotropies. Intristingly, the effect 
of energy loss does not depend on the value of $\left . dE/dx\right |_0$. Note that the total 
deposited momentum for both tested values is 
the same, only the rate of deposition is changed. 
Simulations with only hot spots added in the initial state, which contain the same amount 
of energy as hard partons, but no momentum, lead just to a half of the effect of parton energy
loss.  

\begin{figure}[h]
\centerline{\includegraphics[width=0.65\textwidth]{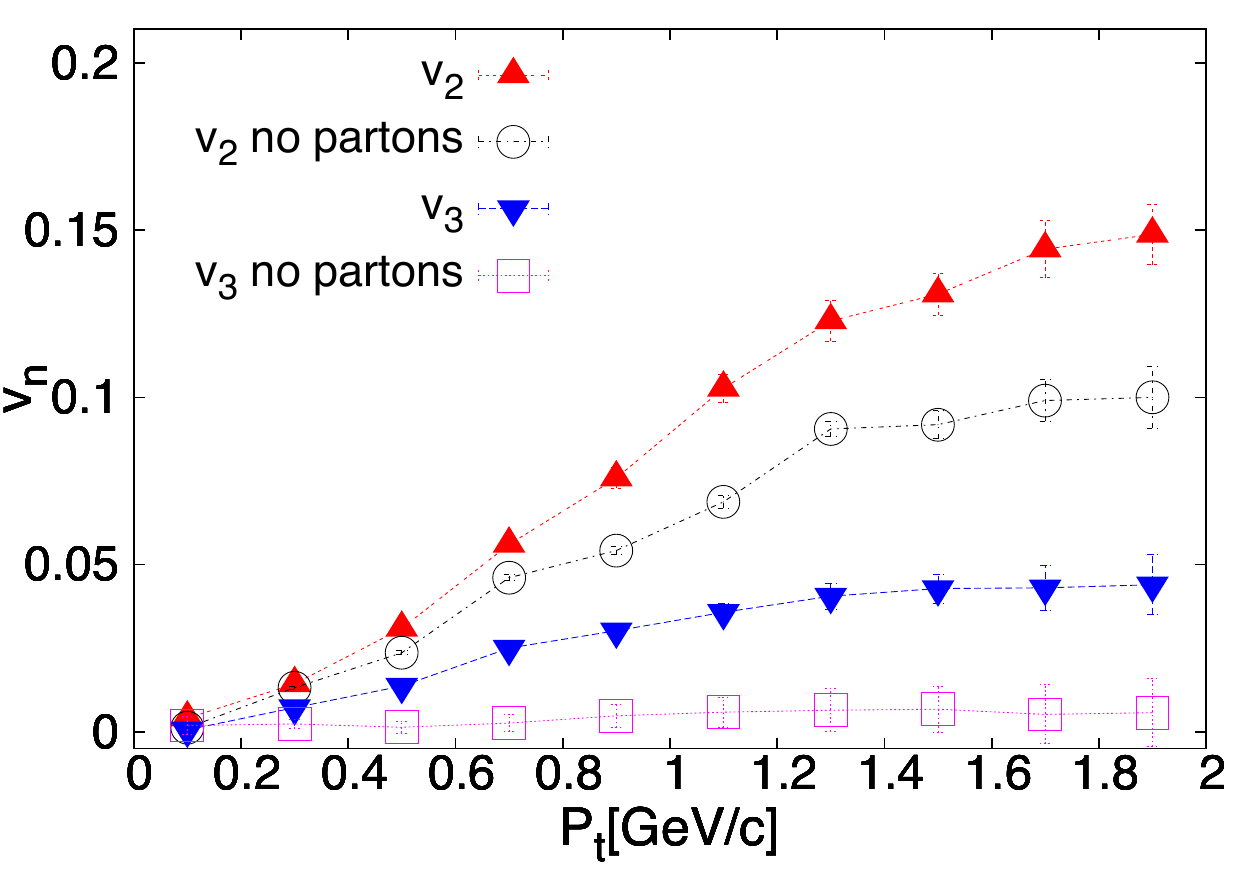}}
\caption{%
Coefficients $v_2$ and $v_3$ from  collisions at impact parameter $b = 6$~fm. 
We show results from simulations with only smooth initial energy density profile
(legend: no partons) as well as with energy loss from hard partons included.  
}%
\label{f:noncen}
\end{figure}

We should prove yet, that the effect is correlated with event geometry so that in non-central 
collisions we indeed obtain positive contribution to the elliptic flow as argued above.
To this end, events corresponding to impact parameter $b = 6$~fm (falls into the 
centrality class 30--40\%) were simulated both with and without the hard partons 
contribution. Main results are summarised in Fig.~\ref{f:noncen}, where we show 
$v_2$ and $v_3$ as functions of $p_t$. 
In non-central collisions, elliptic flow is generated due to the anisotropic matter 
distribution from which the fluid evolves. Hard parton contribution increases the 
observed anisotropy by about 50\%. As for $v_3$, it is not present in case of 
smooth initial conditions due to missing third-order spatial anisotropy is 
solely generated from the hard parton momentum deposition. 

These results indicate that the contribution from hard parton momentum loss is significant 
and should be taken into account in hydrodynamic simulations that aim at the extraction 
of transport properties of the hot and dense nuclear matter. 

An open question is why changing the rate of energy loss has no influence on the generated 
elliptic flow. We speculate that the reason may be that practically all of the 
energy and momentum is deposited very early in the densest period of fireball evolution. 
We want to investigate this question more closely in the future. 

A realistic simulation, also to be accomplished in the future, must include viscosity effects and 
fluctuating initial conditions. There we plan to use a newly developed hydrodynamic model
\cite{zuzka}. 

\textbf{Acknowledgements\ }
BT thanks the organisers for creating stimulating atmosphere at a beautiful locations. 
He also thanks Wojtek Broniowski for insightful discussions. 
This work has been supported in parts by  
APVV-0050-11, VEGA 1/0457/12 (Slovakia) and 
M\v{S}MT grant  LG13031 (Czech Republic).

\end{document}